\documentclass[twocolumn,showpacs,preprintnumbers,amsmath,amssymb]{revtex4}

\usepackage{graphicx}
\usepackage{dcolumn}
\usepackage{bm}

\begin{document}


\title{Electrical spin injection and detection in an InAs quantum well}

\affiliation{}

\author{Hyun Cheol Koo}
\affiliation{Nano Device Research Center, Korea Institute of
Science and Technology, Seoul 136-791, Korea}

\author{Hyunjung Yi}
\affiliation{Nano Device Research Center, Korea Institute of
Science and Technology, Seoul 136-791, Korea}

\author{Donghwa Jung}
\affiliation{Department of Physics, Sejong University, Seoul
143-747, Korea} \affiliation{Nano Device Research Center, Korea
Institute of Science and Technology, Seoul 136-791, Korea}

\author{Jae-Beom Ko}
\affiliation{Nano Device Research Center, Korea Institute of
Science and Technology, Seoul 136-791, Korea}

\author{Seon-Gu Huh}
\affiliation{Department of Physics, Sejong University, Seoul
143-747, Korea} \affiliation{Nano Device Research Center, Korea
Institute of Science and Technology, Seoul 136-791, Korea}

\author{Jonghwa Eom}
\email{eom@sejong.ac.kr} \affiliation{Department of Physics,
Sejong University, Seoul 143-747, Korea} \affiliation{Nano Device
Research Center, Korea Institute of Science and Technology, Seoul
136-791, Korea}

\author{Joonyeon Chang}
\email{presto@kist.re.kr} \affiliation{Nano Device Research
Center, Korea Institute of Science and Technology, Seoul 136-791,
Korea}

\author{Suk-Hee Han}
\affiliation{Nano Device Research Center, Korea Institute of
Science and Technology, Seoul 136-791, Korea}


\date{\today}

\begin{abstract}
We demonstrate fully electrical detection of spin injection in
InAs quantum wells. A spin polarized current is injected from a
Ni$_{81}$Fe$_{19}$ thin film to a two-dimensional electron gas
(2DEG) made of InAs based epitaxial multi-layers. Injected spins
accumulate and diffuse out in the 2DEG, and the spins are
electrically detected by a neighboring Ni$_{81}$Fe$_{19}$
electrode. The observed spin diffusion length is 1.8 $\mu$m at 20
K. The injected spin polarization across the
Ni$_{81}$Fe$_{19}$/InAs interface is 1.9\% at 20 K and remains at
1.4\% even at room temperature.  Our experimental results will
contribute significantly to the realization of a practical spin
field effect transistor.
\end{abstract}


\maketitle

Spintronics is a fascinating new paradigm with the potential to
overcome some of the physical limitations of conventional
electronics. An essential ingredient of spintronic devices is the
presence of a spin-polarized current, which is often generated by
current injection from ferromagnetic metals. However, contemporary
spintronics faces the challenge of developing efficient injection
and detection methods for spin-polarized currents in
semiconductors. In particular, fully electrical spin injection and
detection are the primary prerequisites for realizing a spintronic
device which is compatible with other electronic charge based
devices. Here we provide a manifest evidence of the purely
electrical detection of spin injection and accumulation in InAs
quantum wells. Spins injected from a Ni$_{81}$Fe$_{19}$ thin film
accumulate and diffuse out in the InAs quantum well, with the
spins being electrically detected by a neighbouring
Ni$_{81}$Fe$_{19}$ electrode. The observed spin signals provide us
with substantive information such as the injected spin
polarization and the spin relaxation time in the InAs quantum well
over a wide temperature range up to room temperature.

We have constructed a mesoscopic semiconductor lateral spin valve
that exhibits fully electrical spin injection and detection. The
device consists of two ferromagnetic electrodes and an InAs
high-electron-mobility transistor (InAs HEMT). The InAs HEMT has
been epitaxially grown by molecular beam epitaxy on a
semi-insulated InP(100) substrate. A quantum well, which functions
as a two-dimensional electron gas channel (2DEG), is present at a
depth of 35.5 nm from the top surface. The carrier density and
mobility of 2DEG are n = $6.3 \times 10^{12}$ ($4.6 \times
10^{12}$) cm$^{-2}$ and $\mu$ = 5,700 (34,700)
cm$^{2}$V$^{-1}$s$^{-1}$ at 295 K (20 K). A scanning electron
micrograph of a representative device is shown in Fig.~1(a). The
essential parts of the device are a single 8-$\mu$m-wide InAs HEMT
channel and two ferromagnetic electrodes that work as the spin
injector and detector. The ferromagnetic electrodes have different
aspect ratios, and hence exhibit distinctive
magnetization-switching fields. A desired configuration of
magnetization orientations can be obtained by applying the
external magnetic field along the easy axis of the ferromagnetic
patterns.

The most delicate step in the device fabrication process is
constructing an appropriate interface between the ferromagnetic
electrode and the InAs HEMT. We found that the electrical
detection of spin injection was stable and reproducible only for
devices with an interface resistance for which $R_{F/SC}A_{J}$ =
20$\sim$50 $\Omega$-$\mu$m$^{2}$ at 20 K, where $A_{J}$ is the
interface area and $R_{F/SC}$ is the resistance of the interface
between the Ni$_{81}$Fe$_{19}$ film and the InAs HEMT. A thickness
of 25.5$\sim$32.5 nm was removed from the top surface by Ar-ion
milling and exposed to dry air to yield a very thin oxide layer,
after which an 80-nm-thick Ni$_{81}$Fe$_{19}$ film was sputtered
from a single target in a magnetic field. Figure~1(b) shows a
schematic diagram of the cross section where the ferromagnetic
electrode was deposited. The distance between the ferromagnetic
layer and the InAs quantum well is 3$\sim$10 nm, as measured by
transmission electron microscopy. This distance was found to be
the most suitable for spin injection and detection in our
experiments.

To examine the characteristics of the interface, we compared
effective resistances $R_{F}$ and $R_{SC}$ of a spin diffusion
length for NiFe and the InAs 2DEG, and the interface resistance
$R_{F/SC}$~\cite{2Johnson87,3Rashba00,4Takhashi03}. Using the
values determined from this experiment, we estimate that $R_{F}$ =
$\rho_{F}\lambda_{F}$/$A_{J}$ $\approx$ $4.0 \times 10^{-5}$
$\Omega$ and $R_{SC}$ = $\rho_{SC}\lambda_{SC}$/$A_{SC}$ $\approx$
21 $\Omega$, where $\rho_{x}$ and $\lambda_{x}$ are the
resistivities and the spin diffusion lengths for NiFe (``F") and
the InAs 2DEG (``SC"), respectively, and $A_{SC}$ is the
cross-sectional area of the InAs 2DEG. We used $\lambda_{F}$ = 4.3
nm, as obtained in a previous spin-valve
experiment~\cite{5Dubois99}. The typical $R_{F/SC}$ required to
observe spin signals in this experiment is 1$\sim$10 $\Omega$,
indicating that the interface resistance is in the range $R_{F}
\ll R_{F/SC} < R_{SC}$, which is neither transparent ($R_{F/SC}
\ll R_{F}$) nor strictly tunnelling-like ($R_{F/SC} \gg R_{SC}$).
It should be noted that the conductance-mismatch
model~\cite{7Schmidt00} developed for the transparent limit is not
applicable to the systems in this letter.

Spin injection can be detected electrically by measuring
correlations of the magnetization of two ferromagnetic electrodes.
Two different measurement geometries are usually employed for
these measurements. The first type of measurement is the
four-probe measurement in local spin-valve geometry, in which
voltage drops associated with current flows between the
ferromagnetic electrodes are measured, yielding the resistance in
parallel ($R_{par}$) and antiparallel ($R_{anti}$) magnetization
configurations. Figure~2(a) shows the magnetoresistance measured
in the local spin-valve geometry. The spin-valve effect can be
estimated by the relative magnetoresistance $\Delta R/R_{par}$ =
($R_{anti} - R_{par}$)/$R_{par}$. When the distance between
ferromagnetic electrodes is much smaller than the spin relaxation
length ($\lambda_{F}$) in the ferromagnetic electrodes, $\Delta
R/R_{par}$ has the maximum value of $\beta^{2}$/($1 - \beta^{2}$),
where $\beta$ is the bulk spin polarization in the ferromagnetic
electrode. However, the distance between the ferromagnetic
electrodes is usually larger than the spin relaxation length in a
conventional lateral spin-valve device, in which case the relative
magnetoresistance effect has a very small magnitude. In our
lateral spin-valve devices the observed $\Delta R/R_{par}$ was as
small as 0.029\% (Fig.~2(a)).

The second type of measurement involves nonlocal geometry, which
is used to obtain the spin diffusion length in a metal or a
semiconductor. In this measurement geometry, the voltage is not
measured where the charge current flows; instead, only the
chemical potential that is sensitive to spin accumulation is
measured using a ferromagnetic detector. Figure~2(b) shows the
nonlocal spin signal as a function of magnetic field applied along
the easy axis of ferromagnetic patterns. The chemical potential in
the parallel (antiparallel) magnetization makes an additive
(subtractive) contribution to the nonlocal voltage. The ranges of
magnetic fields where the resistance dips appear in the nonlocal
spin signal match exactly with the ranges where the resistance
maxima occur in the local spin-valve signal of Fig.~2(a). The
characteristics of the observed nonlocal spin signal do not change
even if the injection and detection electrodes are exchanged.
Since the Hall effect is antisymmetric with respect to the zero
magnetic field, it is unlikely that the signal in Fig.~2(b)
contains any contribution from the Hall effect given that the
baseline resistance is flat and no significant antisymmetric
component was detected. For this reason, we abandoned the
speculation that the Hall voltage may be detected using the
nonlocal geometry.

The spin diffusion length and the injected spin polarization were
estimated from an analysis of the space correlation of the
nonlocal spin signal. Figure~3(a) shows the spatial dependence of
the magnitude of the nonlocal spin signal. The spatial dependence
of the nonlocal spin signal $\Delta R$ = ($R_{par} - R_{anti}$) is
known to follow the exponential-decay formula $\Delta R =
(\eta^{2} R_{s} \lambda_{s} / w) \exp(-L/\lambda_{s})$, where
$\eta$ is the injected spin polarization of the current crossing
the NiFe-InAs interface, $R_{s}$ is the sheet resistance of the
InAs 2DEG, $w$ is the width of the InAs 2DEG, $\lambda_{s}$ is the
spin diffusion length of the InAs 2DEG and $L$ is the
center-to-center distance between the ferromagnetic electrodes.
Fitting the obtained data to the formula yields estimates of
$\lambda_{s} \approx 1.8$ $\mu$m and $\eta \approx$ 1.9\% at 20 K.
The black solid line in Fig.~3(a) represents the fitted curve for
20 K, and similar fits were obtained for other temperatures. We
estimate that $\lambda_{s} \approx$ 1.9 $\mu$m and $\eta \approx$
1.7\% at 50 K, $\lambda_{s} \approx$ 1.5 $\mu$m and $\eta \approx$
1.7\% at 100 K, and $\lambda_{s} \approx$ 1.3 $\mu$m and $\eta
\approx$ 1.4\% at 295 K. The injected spin polarization shows a
very weak dependence on temperature (Fig.~3(b)).  These data
represent the determination of $\eta$ purely by electrical
detection of spin injection in a semiconductor. Moreover, the
injected spin polarization remains at 1.4\% up to room
temperature. This weak temperature dependence of $\eta$ is
consistent with a previous study in which the spin injection into
a GaAs quantum well was examined by observing circularly polarized
light~\cite{17Hanbicki02}.

The spin relaxation time can be estimated from $\tau_{s} =
{\lambda_{s}}^{2}/D$. The diffusion constant is given by $D =
1/[N(E_{F})e^{2}R_{s}]$, where $N(E_{F})$ is the density of states
at the Fermi level and $e$ is the electron charge. Figure~3(c)
shows that the temperature dependence of $\tau_{s}$ in the InAs
2DEG is rather weak. In a previous investigation of spin
relaxation using pumping-probe optical orientation spectroscopy,
the temperature dependence of spin relaxation in GaAs-based
quantum wells was observed to depend on the confinement
energy~\cite{18Malinowski}. That study showed that $\tau_{s}$
depends very weakly on temperature for narrow quantum wells (i.e.
with a width of less than 10 nm) while it follows an
inverse-square dependence with temperature for wide quantum wells.
The width of our InAs quantum well was narrow (2 nm), so the weak
temperature dependence of $\tau_{s}$ in Fig.~3(c) is consistent
with spin relaxation in GaAs-based quantum wells. The temperature
dependence of spin relaxation is also consistent with predictions
of the D'yakonov-Kachorovski~\cite{19Dyakonov86} and
D'yakonov-Perel~\cite{20Dyakonov72} mechanisms that the spin
relaxation rate 1/$\tau_{s}$ scales as $T^{3} \tau_{p}$ in a bulk
semiconductor and a wide quantum well, and as $T{E_1}^{2}\tau_{p}$
in a narrow quantum well, where $E_{1}$ is the confinement energy
of the quantum well, $\tau_{p}$ is the momentum scattering time
and $T$ is the temperature. Therefore, $\tau_{s}$ of the narrow
quantum well (as in our InAs HEMT) is expected to be independent
of $T$ when $\tau_{p}$ is inversely proportional to $T$. For
comparison we have also estimated $\tau_{p}$ of the InAs HEMT from
measurements of the sheet resistance and the Hall effect. In the
temperature range of our experiment, $\tau_{p}$ was roughly
inversely proportional to $T$ (Fig.~3(c)).

Purely electrical spin injection and detection in a semiconductor
are important prerequisites for constructing a spin-FET. In the
model spin-FET proposed by Datta and Das~\cite{1Datta90}, spins
injected from a ferromagnetic electrode reach the other
ferromagnetic electrode after precession due to the Rashba
interaction that occurs in a semiconductor channel. Since the
strength of the Rashba spin-orbit interaction is modulated by the
gate electrode in the spin-FET, the output signal is modified by
changes in the spin precession angle. Realizing a spin-FET
requires a semiconductor with a controllable $\alpha$. Our layer
structure is very similar to a previously reported inverted
InGaAs/InAlAs heterostructure with controllable $\alpha$ by a gate
voltage~\cite{24Nitta78}. Consequently, the results in this letter
will contribute significantly to the realization of a practical
spin-FET.

This work was supported by the KIST Vision 21 Program.

\newpage

\baselineskip = 2\baselineskip  

\begin{center}
\begin{figure}
\includegraphics[width=6cm]{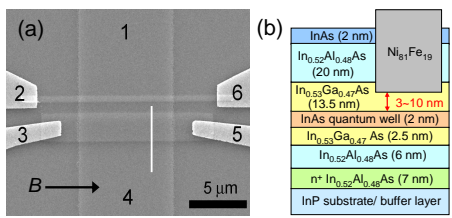}
\caption{(a) Scanning electron micrograph of the mesoscopic
ferromagnet-semiconductor spin-valve device. An 8-$\mu$m-wide
two-dimensional electron gas channel, which is connected by
terminals 1 and 4, is defined by dry etching. The external
magnetic field, $B$, is applied along the easy axis of the
ferromagnetic film. The white line represents a cut line for the
cross-sectional view. (b) Schematic of the cross section where the
ferromagnetic electrode was deposited. The buffer layer consists
of 300-nm-thick In$_{0.52}$Al$_{0.48}$As on top of InP substrate.}
\end{figure}
\end{center}

\begin{center}
\begin{figure}
\includegraphics[width=8cm]{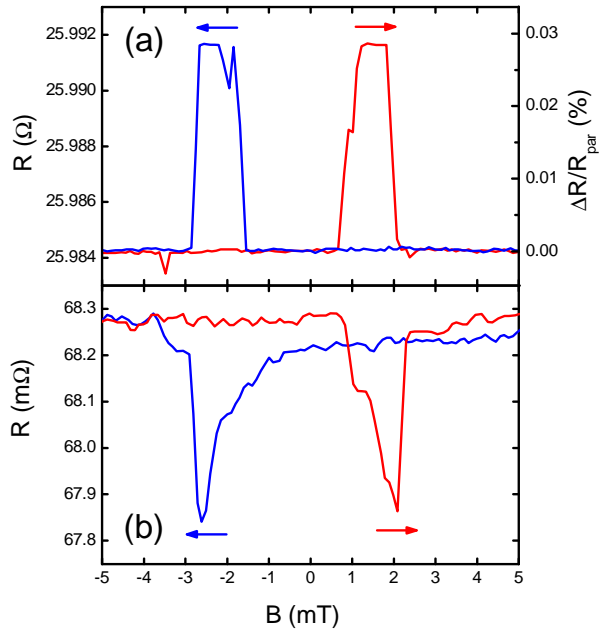}
\caption{(a) Magnetoresistance measured in the local spin-valve
geometry at 20 K. When the current flows from terminals 2 to 3,
the voltage is measured between terminals 6 and 5. The arrows
represent the sweep direction of the magnetic field for each
trace. (b) Nonlocal spin signal at 20 K. The nonlocal voltage is
measured between terminals 3 and 4, while the current flows from
terminal 2 to 1. The data were taken for the device of $L$ = 2.2
$\mu$m. For this device, the distance between the ferromagnetic
layer and the InAs quantum well is $\sim$3 nm.}
\end{figure}
\end{center}

\begin{center}
\begin{figure}
\includegraphics[width=8cm]{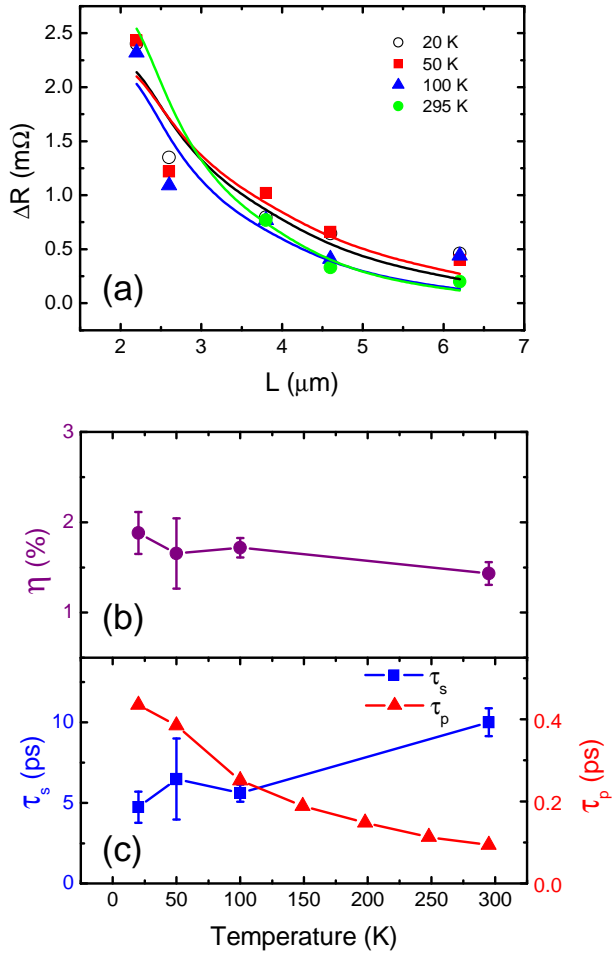}
\caption{(a) Spatial dependence of the magnitude of the nonlocal
spin signal. (b) Temperature dependence of the injected spin
polarization, $\eta$. (c) Temperature dependence of spin
relaxation time ($\tau_{s}$) and momentum scattering time
($\tau_{p}$). The error bars indicate the standard deviations in
our measurements. The distance between the ferromagnetic layer and
the InAs quantum well of all the devices is $\sim$7 nm.}
\end{figure}
\end{center}



\end{document}